# Role of Phonons in Li-Diffusion, Thermal Expansion and Phase Transitions in β-eucryptite: Inelastic Neutron Scattering and Lattice Dynamics Studies

Baltej Singh[a], Mayanak Kumar Gupta[a], Ranjan Mittal[a], Mohamed Zbiri[b], Stephane Rols[b], Sadequa Jahedkhan Patwe[c], Srungarpu Nagabhusan Achary[c], Helmut Schober[b], Avesh Kumar Tyagi[c] and Samrath Lal Chaplot[a] β-eucryptite ($LiAlSiO_4$) shows one-dimensional super-ionic conductivity as well as anisotropic thermal expansion behavior. We have performed inelastic neutron scattering measurements in β-eucryptite over 300-900 K and calculated the phonon spectrum using ab–initio density functional theory method. The calculated energy profile for cooperative lithium ion displacements indicates preferential movement of Li ion along the hexagonal c-axis in the high temperature phase. However, the energy barrier for Li ion diffusion is significantly reduced when a Schottky defect is introduced in the crystal. Further, the anisotropic stress dependence of the phonon spectrum is calculated to obtain the thermal expansion behavior along various axes. The calculations show that the Grüneisen parameters of the low-energy phonon modes around 10 meV have large negative values and govern the negative thermal expansion behavior both along the a- and c- axes. On the other hand, anisotropic elasticity along with anisotropic positive values of the Grüneisen parameters of the high-energy modes in the range 30 to 70 meV are responsible for positive thermal expansion in the a-b plane and negative expansion along the c-axis. The analysis of the polarization vectors of the phonon modes sheds light on the mechanism of the anomalous thermal expansion behavior. We extend the study to discuss the relationship of the soft phonons in the Brillouin zone with the observed high-pressure and high-temperature phase transitions as reported in the literature.

[a]*Solid State Physics Division, Bhabha Atomic Research Centre, Mumbai, 400085, India*
[b]*Institut Laue-Langevin, BP 156, 38042 Grenoble Cedex 9, France*
[c]*Chemistry Division, Bhabha Atomic Research Centre, Mumbai, 400085, India*

## I. INTRODUCTION

Sustained energy demand and the search for alternatives to fossil resources have triggered extensive research for energy storage and in particular battery materials[1-6]. Several options for positive electrodes have been explored in the last few decades, motivated by safety concerns and cost. The field is highly dominated by Li-based compounds[7-13]. The high mobility of Li and the high energy density of these compounds make them ideal candidates for battery materials[14-20]. The dimensionality and topology of the compounds is found to play an important role in the diffusion of Li ions[21, 22]. The layered compounds are found to exhibit particularly high diffusion coefficients for Li-ion[1, 23, 24]. β-eucryptite (LiAlSiO$_4$) finds application as a battery material[25-29] as well as a humidity sensor. It is used in the manufacturing of domestic cookware, high-precision machines and optical devices, solid electrolytes for Li-ion batteries and pressure sensitive switches. The compound has been widely used as low thermal expansion[30-32] fillers in many areas like white ware bodies. Further, high strength and good thermal shock resistance make it a good candidate for applications like heat exchangersin industry.

At room temperature, β-eucryptite crystallizes (Fig. 1) in the space group P6$_4$22.Ithas 84 atoms in a unit cell[25, 33-35],andis basically a stuffed derivative of β-quartz[33]. It can be derived from quartz (SiO$_2$) by replacing half of Si$^{+4}$ with a pair of Al$^{+3}$ and Li$^+$ to keep the charge neutrality of the structure[33]. β-eucryptite contains double helices of SiO$_4$ and AlO$_4$ tetrahedra[35, 36]. This produces alternating layers of Al and Si normal to the hexagonal c-axis, resulting in the doubling of the c-lattice parameter compared to that of quartz[33-35]. Li is present in one-dimensional channels parallel to the c-axis. There are two types of channels (S & A) for Li, with six available tetrahedral sites in each channel. In the channels Li atoms occupy only three of the available sites in an alternating sequence with vacancy sites. There are three secondary (S) and one primary (A) channel. The Li atom in the secondary channels is translated by c/3 with respect to that in the primary channel[37]. So, the Li ions form a superstructure occupying sites in the Si and Al planes with a ratio of 3:1. Li tetrahedra are found to be strongly distorted as revealed by neutron scattering and reverse Monte Carlo modeling[38].

β-eucryptite shows a strong anisotropic thermal expansion behavior[32, 37, 39] ($\alpha_c \approx -2\alpha_a$) with linear thermal expansion coefficients[31, 32, 39] of about $\alpha_a = 8.6 \times 10^{-6}$K$^{-1}$ and $\alpha_c = -18.4 \times 10^{-6}$K$^{-1}$. This results in a very low volume thermal expansion (~1×10$^{-6}$K$^{-1}$) over a wide temperature range of 300-1400 K. This anomalous behavior is suggested to occur due to a rigid rotational tilting of the framework tetrahedral, which leads to an expansion in the (001) plane and a contraction along the c-axis, with increasing temperature[32].

The compound shows one-dimensional superionic conductivity and an order-disorder transition near 755 K[40]. The one-dimensional superionic conductivity is found to be due to inter- and intra-channel correlated motions of Li along the c-axis[25, 41, 42]. Single crystal electrochemical transport measurements show a one-dimensional conductivity (T>673K) due to Li ions, which is about 30 times higher along the c-axis than within the a-b plane at about 773K[25, 41]. The intra channel correlation of Li-Li is reported to be much stronger than the inter-channel correlations[43], which corroborates the cooperative motion of Li along the channel in the superionic regime. Quasi-elastic neutron scattering studies[44] of β-eucryptite show jump diffusion along the hexagonal c-direction of Li, by c/3. A strong inter- and intra-channel correlation is observed upto 1073K[44].

The high-temperature structure modification (Fig. 1) for β-eucryptite was studied by various techniques in the recent years. A lot of discrepancies are reported regarding the structure, temperature and mechanism for the phase transition. A reversible iso-structural phase is formed by Li disordering and geometrical changes in the framework through Al/Si tetrahedral deformation and tilting. The ''average'' structure of the high-temperature phase has three formula units (~21 atoms in the unit cell), with all the Li atoms lying in the Si layers[37]. A first-order framework phase transition is suggested at 673 K with the quartz like high-temperature phase. The high-temperature form is accompanied by a statistical distribution of Li over all the tetrahedral sites of the main channel[36]. Single crystal X- ray diffraction studies show no first order transition. The high-temperature phase is rather observed from the disappearance of superstructure reflections above 733 K due to a combination of normal thermal vibrations of the framework and an induced order-disorder behavior of the Li atoms in the channel[37]. Infrared and Raman spectroscopy[45] show complete positional disordering of Li$^+$ along the structural channels, parallel to the c-axis upon heating above 715 K. The disordering process appears to induce a framework distortion. Moreover, an intermediate phase is suggested to exist between 715- 780 K. However the detailed structure of this intermediate phase is not clear.

Several neutron diffraction studies[32, 39, 40, 43, 46] have been reported on the high-temperature phase transition in β-eucryptite (LiAlSiO$_4$). A high-resolution powder neutron diffraction[46] study is reported as a function of temperature in the range from 573 to 873 K. This study also predicted cooperative motion of Li ions in the compound. The authors have used the split atom model to show that the Li$^+$ ion are hopping between ideal Wyckoff positions and another vacant site, compared to the ideal eucryptite structure. An observed anomaly on cooling at 758 K is attributed to a displacive transition along with ordering of the Li atoms, which is accompanied by formation of an incommensurate structure and doubling of the a-lattice parameter. A second anomaly on further cooling at 698 K is attributed to the disappearance of the incommensurate structure. The co-existence of the commensurate and incommensurate structures is explained by a phenomenological theory based on coupling of charge density waves in these structures[47].

High-pressure studies on β-eucryptite have been performed[48-50] in recent years. Single crystal X-ray and Raman scattering measurements show a pressure-induced transition at 8 kbar and 300K. The high-pressure phase is found to have an elongated a-axis[48] (1.5 times of that of the β-eucryptite). Another high-pressure powder X-ray diffraction measurement[49], at ambient temperature, showed a reversible phase transition from hexagonal β-eucryptite to an orthorhombic phase (ε-eucryptite), between 0.83 and 1.12 GPa. This orthorhombic phase along with β-eucryptite transforms irreversibly to a kinetically hindered hexagonal (α-eucryptite) phase on pressurizing at 2.2 GPa, above 873K[49]. A further increase of the pressure leads to a progressive amorphization at pressures above 4.5 GPa. A completely amorphous state is reached above 17.0 GPa. Reversible β-ε transformation was also observed at 2.0 GPa using in-situ Raman indentations. However, no structural information about the orthorhombic ε-eucryptite was reported[51]. Moreover, recent meta-dynamics simulations reveal the occurrence of an amorphous phase at 3.0 GPa and 300 K which is 19.4% denser than β-eucryptite phase[52].

Classical molecular dynamics simulations[53] were performed to understand the $Li^+$ motion in β-eucryptite. The phonon spectrum of β-eucryptite has been calculated using density functional perturbation method[39, 54]. The authors also reported the calculation of the temperature dependence of the lattice parameters. The density functional theory was used to determine the elastic constants[55]. The calculated elastic constants were found to be consistent with the related measurements[31].

As above introduced, many works have been done using X- ray diffraction, Raman scattering, NMR, neutron diffraction and simulation techniques to understand the thermodynamic behavior of β-eucryptite. However, only a very limited amount of work is available with the aim to addresses the vibrational aspects as a function of temperature and to understand the superionic behavior of the compound. In this study we present the temperature dependent inelastic neutron scattering measurements of β-eucryptite, over the extended temperature range from 313 K to 898 K, to cover the superionic transition as well as the high-temperature structural phase transitions. We have also performed extensive ab-initio lattice dynamical calculations to accompany the measurements; in order to gain deeper insights into the microscopic origins of the anomalous thermal expansion, superionic behavior, and pressure as well as temperature induced phase transitions of the compound. The room temperature and high-temperature phases of β-eucryptite will be labeled below RT and HT, respectively.

## II. EXPERIMENTAL

Polycrystalline samples of β-eucryptite were prepared by solid state reaction of $Li_2CO_3$, $Al_2O_3$ and $SiO_2$. Initially, about 15grams of a homogenous mixture of stoichiometric amounts of reactants (in 1:1:2 molar ratio) are maintained at 800°C for 12 hr and then reground and pressed into a pellet. The pellet is tempered successively at 900°C for 12 h, 1000°C for 12h and then at 1100°C for 12h. After each tempering the pellet is reground and pelletized for the next heat treatment. The bright white crystalline product obtained after the final heat treatment is characterized by powder XRD. The powder XRD data of the final sample show all the peaks corresponding to the β-eucryptite phase (JCPDS-PDF 73-0252).

The inelastic neutron scattering measurements were carried out using the high-flux time-of-flight (IN4C) spectrometer at the Institut Laue Langevin (ILL), France covering a wide range of scattering angles from $10^o$ to $110^o$. Thermal neutrons of wavelength 2.4 Å (14.2 meV) are used for the measurements. The scattering function S(Q, E) is measured in the neutron energy gain mode with a momentum transfer, Q, extending up to 7 $Å^{-1}$. About 10 grams of polycrystalline sample of β-eucryptite have been used for the measurements. The polycrystalline sample of β-eucryptite was put inside a cylindrical niobium sample holder and mounted in a furnace. The sample was heated to 313 K, 498 K, 673 and 898 K, respectively. In order to obtain high-quality data in the entire spectral range, the measurement at 300 K was also performed without the furnace. The data analysis was performed, by averaging the data collected over the angular range of scattering, using ILL software tools[56] to get neutron cross section weighted phonon densities of states. The phonon density of states $g^{(n)}(E)$ in the incoherent[57, 58] one-phonon approximation is extracted as follows

$$g^{(n)}(E) = A \left\langle \frac{e^{2W(Q)}}{Q^2} \frac{E}{n(E,T) + \frac{1}{2} \pm \frac{1}{2}} S(Q,E) \right\rangle$$

$$g^n(E) = B \sum_k \left\{ \frac{4\pi b_k^2}{m_k} \right\} g_k(E)$$

Where $n(E,T) = [\exp(E/k_B T) - 1]^{-1}$, ± represents energy gained/lost by the neutron. $b_k$, $m_k$ and $g_k(E)$ are, respectively, the neutron scattering length, mass, and partial density of states of the $k^{th}$ atom in the unit cell. 'A' and 'B' are normalization constants and 2W (Q) is the Debye-Waller factor. The multiphonon contribution has been calculated using the Sjolander[59] formalism and is subtracted from the room temperature data. The weighting factors $\frac{4\pi b_k^2}{m_k}$ for various atoms in the units of barns/amu are: 0.1974, 0.2645, 0.0557 and 0.0776 for Li, O, Al and Si, respectively. The values of neutron scattering lengths for various atoms can be found from Ref.[60]

## III. COMPUTATIONAL DETAILS

The Vienna based ab-initio simulation package (VASP) wasused for the calculations[61, 62] of the structure and dynamics. The calculations are performed using the projected augmented wave (PAW) formalism of the Kohn-Sham density functional theory within generalized gradient approximation (GGA) for exchange correlation following the parameterization by Perdew, Becke and Ernzerhof.[63, 64]. The plane wave pseudo-potential with a plane wave kinetic energy cutoff of 820 eV was adopted. The integration over the Brillouin zone is sampled using a k-point grid of 2×2×2, generated automatically using the Monkhorst-Pack method[65]. The above parameters were found to be sufficient to obtain a total energy convergence of less than 0.1 meV for the fully relaxed (lattice constants & atomic positions) geometries. The total energy is minimized with respect to structural parameters. The Hellman-Feynman forces are calculated by the finite displacement method (displacement 0.04 Å). Total energies and force calculations are performed for the 162 distinct atomic configurations resulting from symmetrical displacements of the inequivalent atoms along the three Cartesian directions (±x, ±y and ±z). The convergence criteria for the total energy and ionic forces were set to $10^{-8}$ eV and $10^{-5}$ eV Å$^{-1}$. The phonon energies (e.g. dispersion curves, density of states) were extracted from subsequent calculations using the PHONON software[66]. The phonon calculation has been done considering the crystal acoustic sum rule. The slopes of the acoustic branches are used to estimate the elastic constants. Thermal expansion calculation was done using the pressure dependence of phonon frequencies in the entire Brillouin zone. The calculated structures (Table I and II) of the low- and high-temperature phases of β-eucryptite are in agreement with the available experimental data.

## IV. RESULTS AND DISSCUSION

### A. Experimental and Calculated Phonon Spectrum

In order to obtain high-quality data in the entire spectral range (Fig. 2), the measurement was first done at ambient temperature without the sample environment(furnace). This was followed by the temperature-dependent measurements (Fig. 3) from 313 K to 898 K. The comparisons between the measured phonon spectrum at room temperature and the ab-initio calculated one is shown in Fig.2. The general characteristics of the experimental features are very well reproduced by the calculations. The calculated as well as the experimental phonon spectra show that the phonon modes belong to two distinct regions, and exhibit a significant band gap. The first region extends up to 100 meV, composed of the vibrational modes from all the atoms in the unit cell. On the other hand, the high-energy part of the spectrum, between 110-150 meV, corresponds to the internal vibrations of the $SiO_4$ and $AlO_4$ tetrahedra. Moreover the peaks in the calculated spectrum show a tiny downward shift in energy as compared to the inelastic spectrum. This shift is due to the overestimation of the unit cell volume in GGA approximation. The calculated phonon spectra is also compared (Fig. 5) with the Car Parrinello molecular dynamics (CMPD) spectra of the RT phase of β-eucryptite.

The neutron weighted partial density of states (Fig. 2) can also be extracted to identify the atomic contributions in the neutron inelastic spectrum. It can be seen that the major contributions from Li atoms lie below 70 meV, while Si, O and Al atoms contributed in the whole spectral range, up to 140 meV. The partial density of states imply that the double peaks in the experimental spectra, around 10-30 meV, are due to the O atoms. The eigenvector analysis of these modes implies vibrations of the T-O-T (T=Si, Al) bonds, giving rise to a rotational motion of the tetrahedral units as it was also observed in a previous study by Zhang et al[45]. The most intense peak in the spectrum is around 40 meV. It shows maximum contributions from the Li and O atom (Fig.2). This peak is associated with the $LiO_4$ tetrahedral stretching modes, showing isotope shifts of 8.7 cm$^{-1}$(1.08 meV) to 10.6 cm$^{-1}$(1.31 meV), respectively, through $^6$Li/$^7$Li isotope substitution[67]. The eigenvectors of these modes confirmed the vibrational motion of Li atoms along the hexagonal c-axis. The next peak in the spectrum around 50-60 meV originates mainly from the O atoms. This frequency domain is associated with the stretching and bending of Si-O-Al linkages[45], in corner shared $AlO_4$ and $SiO_4$ tetrahedra as the direct Si-O-Si linkage is not favored in β-eucryptite due to the Al avoidance principle[68]. The humps in the experimental spectrum around 65 - 75 meV are due to the Li contribution. These phonon modes correspond to the Si-O-Li bridge vibrations as reported in previous studiesperformed in the range 500-600 cm$^{-1}$ (62-74.5meV)[67].

The maximum contribution from Al and Si is seen from around 70 meV and onward, which ultimately forms a double peak in the experimental spectrum at around 75-95 meV. These types of vibrations have also contributions from the O atoms. The related eigenvectors indicate that the modes originate from vibrations of Al and Si tetrahedra. The high-energy spectrum at 110-140 meV are largely dominated by Si-O and Al-O stretching modes. However, the contribution from Al is relatively small (Fig.2), as compared to that of Si. This indicates that the highest energy peak is mainly governed by $SiO_4$ inter-tetrahedral vibrational modes with a small component from $AlO_4$ tetrahedral modes, in accordance with previous spectroscopic studies[45]. Dispersive nature of the stretching mode spectrum is attributed to anisotropic bonding in $SiO_4$ and $AlO_4$ polyhedra.

**B. Temperature Dependence of the Phonon Spectrum**

At 313 K, the large background from the furnace prevented obtaining the full-energy range phonon spectra. Upon heating, the high-energy phonon modes get populated, and therefore data can map the full energy range, up to 150 meV. The measurements are carried out at four different temperatures namely 313 K, 498 K, 673 K and 898 K, across the order-disorder phase transition reported at 698 K. The data are corrected for multiphonon contribution as calculated using Sjolander[59] formalism. The temperature-dependent spectra consist of several peaks around 15, 25, 40, 55, 85 and 125 meV. A previous infrared study of β-$LiAlSiO_4$ showed absorption peaks in the energy range of 100 - 1200 $cm^{-1}$ (12 to 150 meV)[45]. The temperature dependent phonon density of states show (Fig. 3) that Li-based features (35-45 and 60-65 meV) are broadened upon heating. This might be due to the increasing mean square amplitude of Li atoms as a function of temperature. The peak about 50 meV in the measured spectra also shifts towards low energies, with increasing temperature. The modes about 50 meV involve bending of $AlO_4$ and $SiO_4$ polyhedra connected through Al-O-Si linkage. The increase in the unit cell volume will result in a slight increase of the Al-O and Si-O bonds, forming the Al-O-Si linkage. This would in turn slightly decrease the energy of these modes upon heating. The observed temperature-dependent red-shift in the high-energy feature of the spectra (110- 140 meV) is due to the softening of the Si-O and Al-O stretching modes of the $AlO_4$ and $SiO_4$ polyhedral units.

**C. Phonon Spectrum in High Temperature Phase of β-eucryptite**

The high temperature phase is suggested to have Li positional disordering[40, 46]. Its unit cell (in the space group $P6_222$) is one fourth of the room temperature cell and contains 21 atoms, i.e. 3 formula units per unit cell[37, 40]. The phonon spectrum of the high temperature phase has been calculated (Fig.

4) assuming an ordered structure[37], as previously done in other works[39, 54]. This spectrum is compared (Fig. 5) with linear response (LR) density functional theory (DFT) calculations from the literature. The calculated partial phonon density of states (Fig. 4) shows that the energy ranges of the contributions due to the Li atom in the HT and RT phases is almost the same. As noted above, in the RT phase we have two channels for movement of Li atoms and they become equivalent in the HT-phase. The difference is also reflected in the calculated partial density of states of Li atoms. The two peak structure around 60 meV in the room temperature phase evolves into a single peak in the HT-Phase (Fig. 4).

The calculated values of the mean-squared amplitude ($u^2$) in the HT and LT phases show that the Li atoms have much larger vibrational amplitudes along the c-axis than in the ab-plane (Fig. 7). This points towards an easy channel movement of Li atoms along the c-axis, and providing a maximum conduction probability. This also matches the experimental observations that superionic conduction of Li ions in the material is along the c-axis. The c-component of the $u^2$ in the HT-phase is found to be closely similar to the RT phase.

The phase transition from the HT to the RT phase involves almost exact doubling of the a-lattice parameter and a slight decrease in the c-lattice parameter. Therefore, the transition from the HT to the RT-phase has been argued[46] to be due to softening of the M-point phonon mode in the HT-phase. In order to estimate the effect of the change in the c-lattice parameter on the phonon spectra we have performed calculations in the HT phase at c-lattice parameters of 11.476 Å and 11.337 Å (which corresponds to the change in going to the RT phase). The comparison between the calculated phonon dispersion relations along the high symmetry directions in the Brillouin zone of the hexagonal unit cell at these two volumes is shown in Fig. 8. We find that on decrease of the c-lattice parameter, several phonons at the Γ, M, A and L points are found to soften. Since the Li-positional disorder is not included in the calculations, it appears that the disorder has an important role in the HT-RT phase transition.

## D. Energy Barrier for Li Diffusion

β-eucryptite shows one-dimensional superionic conductivity (upto 875K). The compound is found to undergo the order-disorder transition at 698 K followed by the framework displacive transition on further heating at 758 K[46]. The room temperature structure of β-eucryptite contains three different sites of Li atoms (Li1, Li2 and Li3). The Li atoms are fourfold coordinated in either the S or A channels. In the A channel three Li (Li1) are coplanar with Al tetrahedral planes, whereas in the S channel Li (Li2 & Li3) atoms are coplanar with Si tetrahedral planes. In a single unit cell, the ratio Li$_S$:

Li$_A$ of 3:1 is energetically favorable[39]. The environment around Li atoms in both channels is different. There is no noticeable cross-channel motion of Li atoms up to 875 K[46].

The calculation of the barrier height for the cooperative motion of Li diffusion has been reported[39]. However no distinction has been made for diffusion of Li in S or A channels. It might be possible that the energy profile for diffusion of lithium may be favorable in one of the two channels. Here we have calculated (Fig. 9) the energy profile for diffusion of Li atoms in various channels. The structure corresponding to 300 K has been used in the calculations. It may be noted that the depth of the secondary minima in the energy profile for the channel 'A' is deeper (0.12 eV/Li-atom) in comparison to that of channel S (0.08 eV/Li-atom), indicating that the probability of Li in the channel 'A' to stay in a secondary minimum is higher than for channel 'S'. This confirms that, provided conductive conditions prevail, the Li atoms in channel 'A' might move. At high temperatures, above 758 K, the availability of defects and vacancies could lead to a simultaneous movement of several lithium ions towards their nearest vacant site. This could lead to a macroscopic effect of diffusion, which is observed in β-eucryptite above 698K.

In the HT-phase all the channels containing Li are equivalent and there are six available sites for Li in each channel[40]. The energy profile for Li channel diffusion has been calculated. We find that the height of energy barrier for Li channel diffusion along the c-axis in the HT phase is smaller than that in the RT (S & A channel) phase by an amount equal to 38 meV. So there exists a large probability for Li to be present among all the available sites in the HT phase. This indicates that the HT-phase has easier one-dimensional conduction of Li.

Moreover, there is a high probability of occurrence of defects at elevated temperatures. We have also calculated the barrier height for the Li diffusion by considering a Schottky vacancy (Li vacancy). A Li vacancy is created at (0,0,c/3). In order to study the barrier height for Li vacancy motion along the hexagonal c-axis, another Li atom is moved in steps to the vacancy. This has resulted in an energy profile for the lithium atom. The comparison of energy profile for Li atom motion for the single Li vacancy and the intra channel correlated motion of Li is shown in Fig. 14. It can be seen that the diffusion of Li atoms is favorable for a structure possessing Li vacancies, compared to the case with a cooperative motion of Li atoms in a perfect crystal. The results are in agreement with the experimental observations[25, 27, 28, 41, 44, 46, 69].

## E. Elastic constants and Anomalous Thermal Expansion Behaviour

The elastic constants(III) are calculated using the symmetry-general least square method[70] as implemented in VASP5.2. The values are derived from the strain−stress relationships obtained from six finite distortions of the equilibrium lattice. For small deformations we remain in the elastic domain of the solid and a quadratic dependence of the total energy with respect to the strain is expected (Hooke's law). There are 5 linearly independent elastic constants[71] for a hexagonal cell of β-eucryptite namely $c_{11}$, $c_{12}$, $c_{13}$, $c_{33}$ and $c_{44}$. The sixth elastic constant, $c_{66}$ is a linear combination of $c_{11}$ and $c_{12}$, $c_{66} = (c_{11} - c_{12})/2$. The bulk modulus has been calculated using the relation,

$$B = \frac{c_{33}(c_{11} + c_{12}) - 2c_{13}^2}{(c_{11} + c_{12}) + 2c_{33} - 4c_{13}}$$

The calculated (Table III) elastic constants and bulk modulus are in good agreement with the reported experimental data as well as other estimations from the literature.

The linear thermal expansion coefficients along the 'a' and 'c' -axes have been calculated within the quasiharmonic approximation[72]. The calculations require anisotropic pressure dependence of phonon energies in the entire Brillouin zone. The phonon energies in the entire Brillouin zone are calculated at ambient pressure. Ananisotropic stress of 2.5 kbar is implemented by changing the lattice constant 'a' and keeping the 'c' parameter constant; and vice versa. These calculations are subsequently used to obtain the mode Grüneisen parameters using the relation[72],

$$\Gamma_l(E_{q,v}) = -\frac{\partial ln E_{q,v}}{\partial ln l}; \; l = a, b, c$$

Where $E_{q,v}$ is the energy of $v^{th}$ phonon mode at point q in the Brillouin zone. In hexagonal system, Grüneisen parameters $\Gamma_a = \Gamma_b$.

The anisotropic linear thermal expansion coefficients are given by[73]:

$$\alpha_a(T) = \frac{1}{V_0} \sum_{q,v} C_v(q, v, T) \left[ s_{11}\Gamma_a + s_{12}\Gamma_b + s_{13}\Gamma_c \right]$$

$$\alpha_c(T) = \frac{1}{V_0} \sum_{q,v} C_v(q,v,T) \left[ s_{31}\Gamma_a + s_{32}\Gamma_b + s_{33}\Gamma_c \right]$$

Where $s_{ij}$ are elements (Table IV) of elastic compliances matrix $S=C^{-1}$, $V_0$ is volume at ambient conditions and $C_v(q,v,T)$ is the specific heat at constant volume for $v^{th}$ phonon mode at point **q** in the Brillouin zone given as

$$C_v(q,v,T) = E_{q,v} \times \frac{\partial}{\partial T} \left[ \exp\left(\frac{E_{q,v}}{k_B T}\right) - 1 \right]^{-1}$$

The volume thermal expansion coefficient for hexagonal system is given by:

$$\alpha_V = (2\alpha_a + \alpha_c)$$

The calculated temperature dependence of the lattice parameters is in excellent agreement (Fig. 10) with the experimental data[32, 39]. The calculated anisotropic linear thermal expansion coefficients at 300 K are $\alpha_a = 6.0 \times 10^{-6}$ K$^{-1}$ and $\alpha_c = -13.7 \times 10^{-6}$ K$^{-1}$. They compare very well with available experimental values[32] of $\alpha_a = 7.26 \times 10^{-6}$ K$^{-1}$ and $\alpha_c = -16.35 \times 10^{-6}$ K$^{-1}$. The thermal expansion coefficient for the RT-phase in the a-b plane ($\alpha_a(T)$), upto 200 K, has a negative value (Fig. 10(b)). It evolves to a positive value at higher temperatures.

The calculated lattice parameters in the temperature range up to 600 K are in a very good agreement with the available experimental data. However, above 600 K, the calculated parameters deviate slightly (Fig. 10) from the experiments. This might be due to the fact that at high temperature the explicit anharmonicity of phonons plays an important role. This effect has not been considered in the thermal expansion calculation, which includes only the pressure dependence of phonon energies (implicit contribution). It can be seen that the compound shows negative volume thermal expansion behaviour below 300 K, and a positive expansion at higher temperatures. The compound has a very small volume thermal expansion coefficient ($\alpha_V = -1.8 \times 10^{-6}$ K$^{-1}$) at 300 K, which makes it suitable for application as a high thermal shock resistance material.

The partial density of states of various atoms (Fig. 4) is used for the calculation of mean squared displacements of various atoms, $<(u^2)>$, (Fig. 6) arising from all phonons of energy E in the Brillouin zone, as follows:

$$< u_k^2(T) > = \int (n + \frac{1}{2}) \frac{\hbar}{m_k E_{q,v}} g_k(E_{q,v}) dE_{q,v}$$

Where $n = [\exp\left(\frac{E_{q,v}}{k_B T}\right) - 1]^{-1}$, $g_k(E_{q,v})$ and $m_k$ are the atomic partial density of states and mass of the $k^{th}$ atom in the unit cell, respectively. As shown in Fig. 6(a) the equal amplitude of all the atoms up to 6 meV indicates that the modes are acoustic. For energy range 6-20 meV, we find large amplitudes for oxygen atoms in comparison to Al and Si atoms. This indicates that these modes involve a rotation of the $AlO_4$ and $SiO_4$ tetrahedra. Further, above 20 meV, the amplitude of Li atoms is found to be very large as compared to other atoms. This may involve translational motion of Li.

The phonon dispersion relation (Fig 11) in the RT-phase has been calculated along the high symmetry directions of the hexagonal cell at ambient, as well as at 0.5 GPa. It can be seen that some of the zone centre modes are highly anharmonic. The eigenvectors of some of these anharmonic modes are analyzed (Fig. 12) to understand the mechanism of anisotropy in the thermal expansion behavior of the compound.

The Γ-point mode of 7.0 meV (Fig. 12) has Grüneisen parameters values of $\Gamma_a$=-31.0 and $\Gamma_c$=-30.8. This mode shows an anti-phase rotation of adjacent polyhedral units ($AlO_4$ or $SiO_4$) about all the three directions. The central atom (Al or Si) of the polyhedral units has a small displacement in comparison with oxygen atoms. The next Γ-point mode of 9.6 meV ($\Gamma_a$=-13.6, $\Gamma_c$=-15.3) also involves anti-phase rotation of $AlO_4$ and $SiO_4$ but only about b-axis. Here the amplitude of rotation for $SiO_4$ is large in comparison to $AlO_4$ polyhedral units. The Li atoms in both these modes vibrate in the ab-plane.

The high energy, Γ-point, modes of energy 31.7 meV ($\Gamma_a$=2.3, $\Gamma_c$=1.0) and 55.2 meV ($\Gamma_a$=3.6, $\Gamma_c$=1.5) (Fig. 12) have significant positive Grüneisen parameter value. These modes get populated as temperature is increased and might give expansion in the a-b plane at higher temperature. These modes are highly contributed by Li motion with very small component of Al and Si tetrahedral rotation and distortions. The mode at 31.7 meV involve the translational motion of Li along the hexagonal channel and might be correlated with the Li diffusion along c-axis, while the other mode at 55.2 meV, involves the Li translation motion in a-b plane.

The M-Point mode (Fig. 12) of 8.7 meV ($\Gamma_a$=-3.9, $\Gamma_c$=-3.3) reflects a sliding motion of SiO$_4$ tetrahedral units in the a-b plane. The adjacent SiO$_4$ layers move in opposite directions along the b-axis. This gives rise to a rotation of the involved AlO$_4$ polyhedra. The Li atoms in adjacent layers move along b- and c-axis. This type of dynamics induces folding of spirals containing (Al,Si)O$_4$ polyhedra producing contraction in the system. The A-Point mode at 8.3 meV ($\Gamma_a$=-18.0, $\Gamma_c$=-18.9) shows (Fig. 12) AlO$_4$ polyhedral rotation about a-axis while SiO$_4$ polyhedral rotation about b-axis. Here, oxygen atoms connected to the tetrahedral units have different amplitude and may produce polyhedral distortion. The lithium motion lies in the a-b plane.

At low temperature, low-energy modes contribute significantly to the thermal expansion behavior. Although the above mentioned low-energy modes have almost similar values of $\Gamma_a$ and $\Gamma_c$ ($\Gamma_a \approx \Gamma_c$), yet the anisotropy in thermal expansion behavior is there. This anisotropy can be understood in terms of elastic properties of β-eucryptite.

$$\alpha_a \propto [s_{11} + s_{12}]\Gamma_a + s_{13}\Gamma_c = 0.008\Gamma_a - 0.005\Gamma_c$$

and

$$\alpha_c \propto [s_{31} + s_{32}]\Gamma_a + s_{33}\Gamma_c = -0.010\Gamma_a + 0.014\Gamma_c$$

where $[s_{11} + s_{12}] = 0.008$, $[s_{31} + s_{32}] = -0.010$, $s_{13}$= -0.005 and $s_{33}$= 0.014 in GPa$^{-1}$ units (Table IV).

We note that for low-energy modes $\Gamma_a \approx \Gamma_c < 0$, which imply a negative thermal expansion both along a- and c-axis at low temperatures [Fig 10(b)]. However, at high temperatures the entire spectra will contribute to the thermal expansion behaviour and larger positive values of $\Gamma_a$ than those of $\Gamma_c$ result in a positive thermal expansion in *a-b* plane and negative thermal expansion along c-axis.

**F. Temperature Dependence of phonon modes in the RT-Phase**

The low-energy optic mode of 7.0meV ($\Gamma_a$= -31.0, $\Gamma_c$= -30.8) is found to contribute considerably to the large negative thermal expansion behavior. We have calculated the temperature dependence of this mode. The change in phonon energy of a mode originates from the implicit and explicit anharmonicities[74, 75]. The implicit part arises from the unit cell volume dependence of phonon frequencies while the explicit contribution comes from the large thermal amplitude of atoms. We have calculated the explicit contribution due to the anharmonic potential of the phonon mode, neglecting other terms such as phonon – phonon interaction. We have calculated the potential energy profile (Fig. 13) of this mode. The crystal potential energy is then fitted to the expression $V(\xi_j) = a_{0,j} + a_{2,j}\xi_j^2 + a_{3,j}\xi_j^3 + a_{4,j}\xi_j^4$ (where $\xi_j$ is the normal coordinate of the j$^{th}$ phonon mode and $a_{2,j}$, $a_{3,j}$ and $a_{4,j}$ are the coefficients of the harmonic and third and fourth order anharmonic terms respectively). The energy of the modes may increase or decrease with increase of temperature, depending on the sign (negative or positive) of the coefficients of the anharmonic terms. The cubic parameter $a_{3,j}$ is found to be zero. The fitted values of $a_{2,j}$ and $a_{4,j}$ are found to be $6.39 \times 10^{-3}$ and $2.86 \times 10^{-5}$. The details of the perturbation method as used for calculation of the temperature dependence of phonon frequency are given in Ref.[74-76]. The estimated change in energy of the phonon modes upon the increase of temperature from 0 to 500 K is shown in Fig. 14. The mode shows (Fig. 14) a strong explicit temperature dependence. The energy of the mode increases from 7.0 meV at 0 K to about 8.1 meV at 500 K. However the temperature variation of volume thermal expansion shows that it is negative up to 100 meV, then a positive behavior is observed at high temperatures. The mode has a very high negative Grüneisen parameter $\Gamma_a = -31.0$ and $\Gamma_c = -30.8$, indicating a strong sensitivity to the temperature change. The overall temperature dependence (implicit + explicit) shows (Fig. 14) that the mode energy at 500 K is 7.5 meV.

**G. Phonon Softening upon Compression in the RT-Phase**

High pressure X-ray diffraction measurements show that β-eucryptite undergoes a reversible orthorhombic phase (ε-eucryptite) transition[49] between 0.83 and 1.12 GPa, at ambient temperature. Upon increasing the pressure, a progressive amorphization[50] is observed above 4.5 GPa, then a complete amorphization occurs above 17.0 GPa. We have calculated the pressure dependence of the phonon spectrum in the entire Brillouin zone for β-eucryptite. We found that at 2 GPa (Fig.15) one of the zone centre optic phonon modes becomes unstable. The softening of the mode may trigger instability in the crystal lattice and may be responsible for the high-pressure phase transition. The eigenvector of the mode (Fig.12) suggests that it involves a rotation of the $AlO_4$ and $SiO_4$ polyhedral in the b-c plane. We find that the same mode is responsible for the negative thermal expansion behavior along the a- and c-axes. This confirms a fundamental perspective that negative thermal expansion (NTE) and pressure induced transition could both originate from the flexible nature of the structural framework of the crystals. Moreover, the pressure induced amorphization at higher pressures is also believed to be due to the flexible structural framework[77].

## V. CONCLUSIONS

Measurements of the temperature dependence of phonon spectrum together with ab-inito phonon calculations have been successfully used to understand the phonon dynamics in both the RT and HT phases of β-eucryptite. The phonon calculations are a useful tool for the interpretation of the experimental data. We have calculated the energy barrier profile for the diffusion of lithium atoms along the available channels located along the hexagonal c-axis in both the RT and HT phases. This reveals that the so called 'A' channel in the RT phase is a more favorable route than the 'S' channel. In the high-temperature phase, all the channels of the framework are equivalent, and the barrier is reduced considerably. The energy barrier is further lowered when a lithium vacancy is introduced in the crystal. The pressure dependence of the phonon spectrum is used to understand the anisotropic thermal expansion behavior. We find that the high-energy modes involving translational motions of Li result in a positive thermal expansion along the a-axis and negative expansion along the c-axis. Such an anisotropic behavior is a result of anisotropic elasticity and anisotropic Grüneisen parameters. On the other hand, low-energy phonons involving rotational degrees of freedom as well as a distortion of polyhedral are responsible for the negative thermal expansion behavior at low temperature both along a- and c-axis. We find the phonon softening at several points in the Brillouin zone in the HT phase. However, the phase transition from the HT to RT phase in β-eucryptite also involves disorder of the Li-atoms. In the RT-phase, on compression to about 2 GPa, a Γ-point mode softens, which may be related to the high-pressure phase transition. The present work offers a fundamental microscopic framework with the perspective to improve the utilization of the studied material for practical applications.

**Acknowledgements**

S. L. Chaplot would like to thank the Department of Atomic Energy, India for the award of Raja Ramanna Fellowship.

TABLE. I The experimentally refined and calculated structure of β-eucryptite (RT- Phase). The atomic positions in the space group P6$_4$22 are Li1 at *3b(x,y,z)*, Li2 at *3c(x,y,z)*, Li3 at *6f(x,y,z)*, and O1, O2, O3 and O4 at *12k(x,y,z)*, and Al1 at *6h(x,y,z)*, Al2 at *6j(x,y,z)*, and Si1 at *6g(x,y,z)* and Si2 at *6i(x,y,z)* Wyckoff site.

|  |  | Experimental[37] | | | Calculated | | |
|---|---|---|---|---|---|---|---|
| a (Å) |  | 10.497 | | | 10.574 | | |
| c (Å) |  | 11.200 | | | 11.410 | | |
| Atoms | Wyckoff site | x | y | z | x | y | z |
| Li(1) | 3b | 0.000 | 0.000 | 0.500 | 0.000 | 0.000 | 0.500 |
| Li(2) | 3c | 0.500 | 0.000 | 0.000 | 0.500 | 0.000 | 0.000 |
| Li(3) | 6f | 0.500 | 0.000 | 0.324 | 0.500 | 0.000 | 0.328 |
| O(1) | 12k | 0.0853 | 0.1942 | 0.2428 | 0.087 | 0.198 | 0.242 |
| O(2) | 12k | 0.6023 | 0.7008 | 0.2651 | 0.601 | 0.697 | 0.261 |
| O(3) | 12k | 0.1101 | 0.7099 | 0.2597 | 0.106 | 0.706 | 0.263 |
| O(4) | 12k | 0.5902 | 0.2011 | 0.2494 | 0.592 | 0.1998 | 0.251 |
| Al(1) | 6h | 0.2520 | 0.000 | 0.500 | 0.251 | 0.000 | 0.500 |
| Al(2) | 6j | 0.2506 | 0.5012 | 0.500 | 0.251 | 0.502 | 0.500 |
| Si(1) | 6g | 0.2486 | 0.000 | 0.000 | 0.248 | 0.000 | 0.000 |
| Si(2) | 6i | 0.2477 | 0.4954 | 0.000 | 0.248 | 0.496 | 0.000 |

TABLE II. The experimental and calculated structure in the high temperature(HT) phase of β-eucryptite. The 'a' lattice parameter is expected to be about half of that in the (RT)β-eucryptite phase, while c is expected to remain almost unchanged. The unit cell in the space group P6$_2$22 have Li atom at *3a(x,y,z)*, O at *12k(x,y,z)*, Al at *3d(x,y,z)*, and Si at *3c(x,y,z)* Wyckoff site.

|  |  | Experimental[17] | | | Calculated | | |
|---|---|---|---|---|---|---|---|
| a (Å) |  | 5.260(5) | | | 5.288 | | |
| c (Å) |  | 11.095(1) | | | 11.469 | | |
| Atom | Wyckoff site | x | y | z | x | y | z |
| Li | 3a | 0 | 0 | 0 | 0 | 0 | 0 |
| O | 12k | 0.2111 | 0.4046 | 0.2540 | 0.205 | 0.399 | 0.256 |
| Al | 3d | 0.5 | 0 | 0.5 | 0.5 | 0 | 0.5 |
| Si | 3c | 0.5 | 0 | 0 | 0.5 | 0 | 0 |

TABLE III. Calculated and experimental elastic constants in GPa units in the RT and HT phases.

|  | $C_{11}$ | $C_{12}$ | $C_{13}$ | $C_{33}$ | $C_{44}$ | $C_{66}$ | Bulk Modulus |
|---|---|---|---|---|---|---|---|
| Our Calculations, RT | 171.6 | 79.0 | 92.9 | 140.0 | 60.0 | 46.3 | 113.1 |
| Experimental, RT[78] | 176.3 | 68.5 | 89.8 | 139.9 | 61.2 | 53.9 | 109.9 |
| Calculation RT[79] | 178.9 | 102.8 | 118.3 | 181.3 | 47.4 | 38.1 | 134.9 |
| Calculation RT[55] | 165.6 | 71.0 | 78.6 | 132.8 | 58.7 | 47.3 | 101.5 |
| Our Calculation HT | 179.3 | 84.7 | 101.0 | 152.0 | 47.3 | 47.3 | 120.3 |

TABLE IV. Calculated elastic compliance matrix components ($s_{ij}$) in GPa$^{-1}$ units for the RT phases.

| $s_{11}$ | $s_{12}$ | $s_{13}$ | $s_{33}$ | $s_{44}$ | $s_{66}$ |
|---|---|---|---|---|---|
| 0.00932 | -0.00148 | -0.00520 | 0.01405 | 0.01668 | 0.02163 |

FIG. 1 (Color online) Schematic representation of the unit cell structure of (a) room temperature phase of β-eucryptite (b) high temperature phase (supercell 2×2×1) of β-eucryptite, projected along different axes. $AlO_4$ and $SiO_4$ tetrahedra are in purple and reddish brown color respectively.

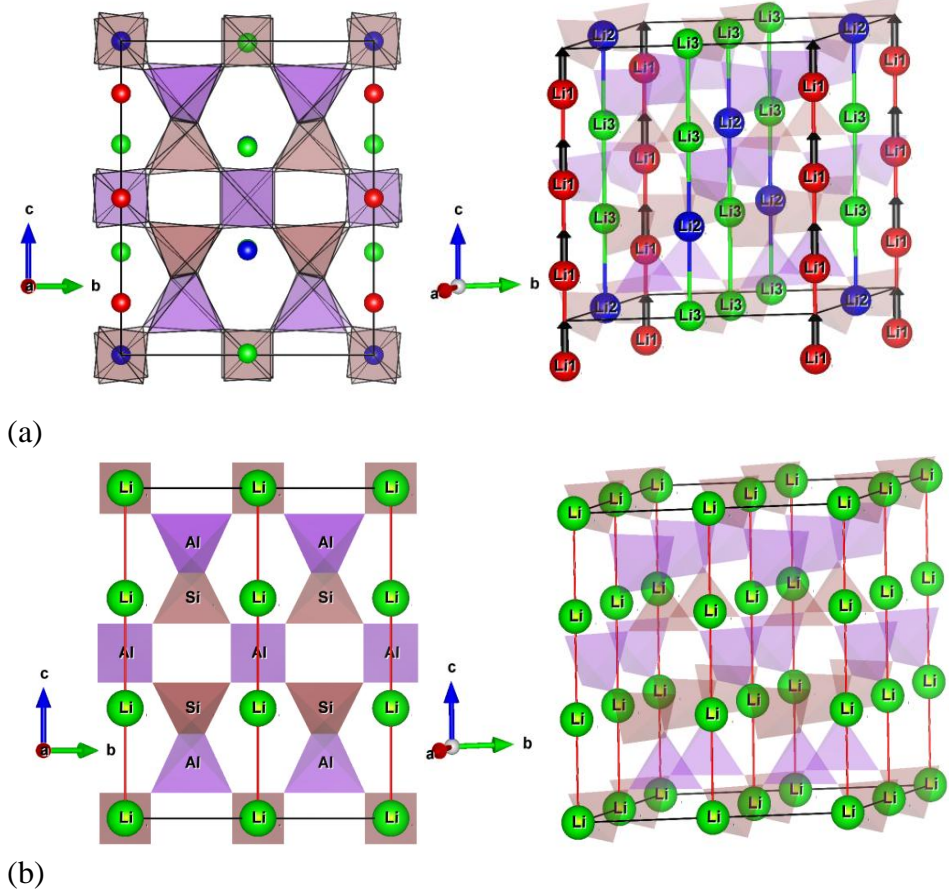

(a)

(b)

FIG. 2 (Color online) Experimental and computed phonon density of states in RT-Phase of β-eucryptite. The calculated partial phonon density of states of various atoms is also shown. In order to obtain high quality data in the entire spectral range, the measurement at 300 K was performed outside the furnace.

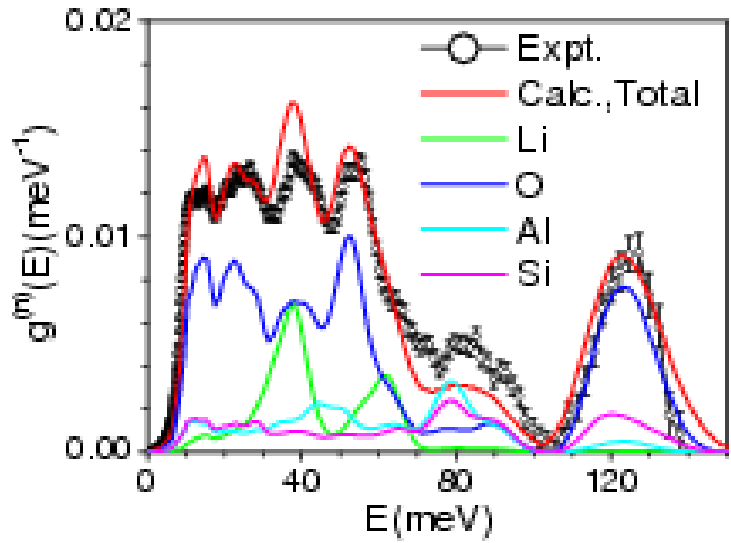

FIG. 3 (Color online) Temperature dependent experimental phonon densities of states. The measurements are carried out using the high temperature furnace available at ILL. For clarity, the spectra are shifted vertically.

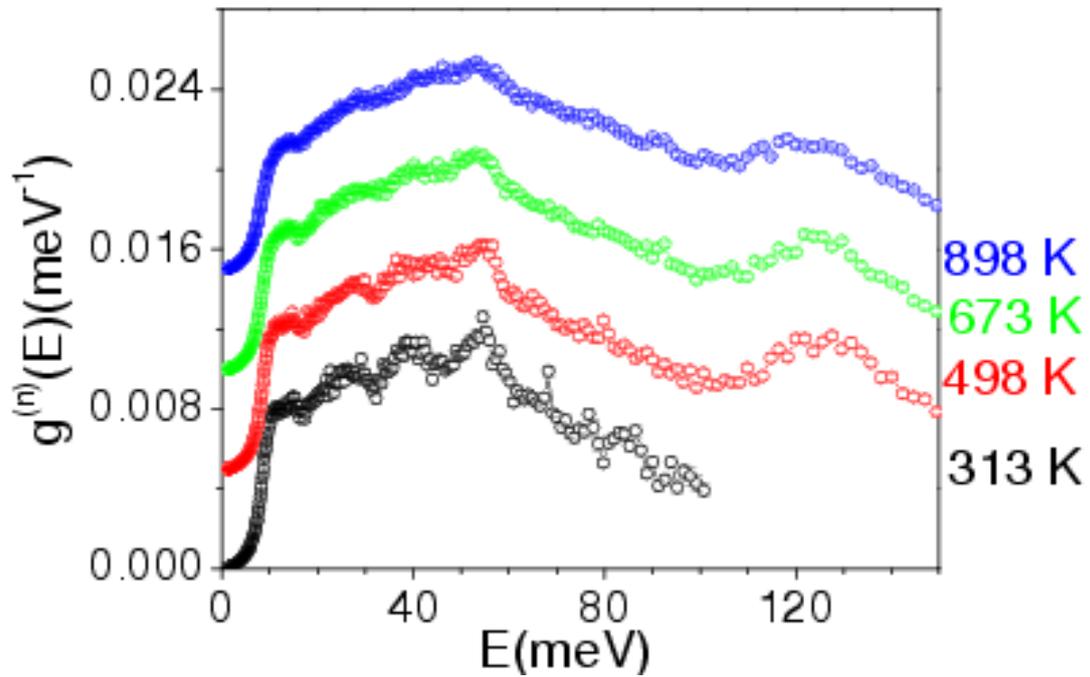

FIG. 4 (Color online): Calculated phonon density of states of various atoms in the RT (solid line) and HT (dashed line) phases of β-eucryptite.

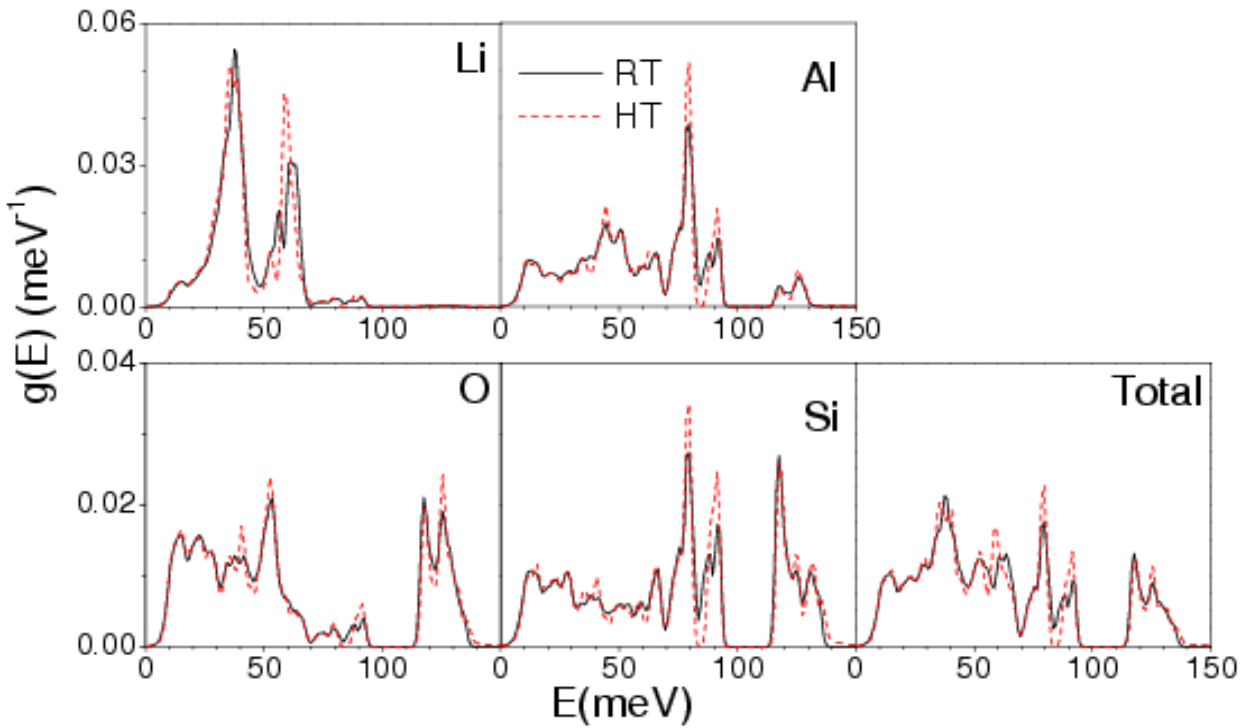

FIG. 5 (Color online): Calculated phonon density of states in (a) HT-Phase and (b) RT-Phase of β-eucryptite, from the present work and from the literature[54].

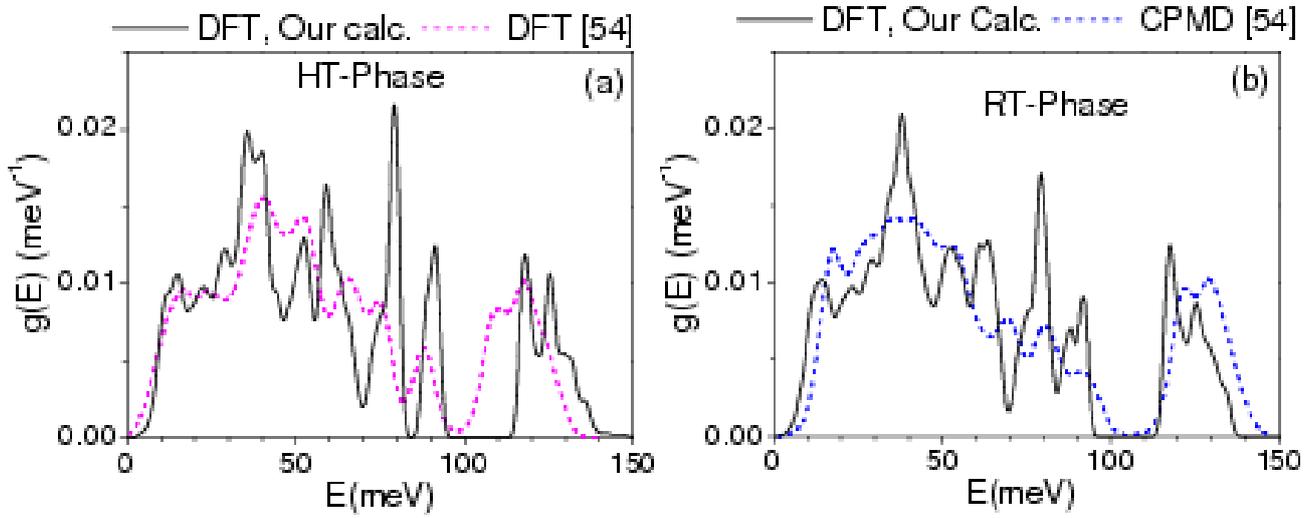

FIG. 6 (Color online) (a) The calculated contribution to the mean squared amplitude of the various atoms in the RT phase of β-eucryptite arising from phonons of energy E at $T$=300 K.(b) The calculated mean square amplitude of different atoms in the RT phase of β-eucryptite as a function of temperature.

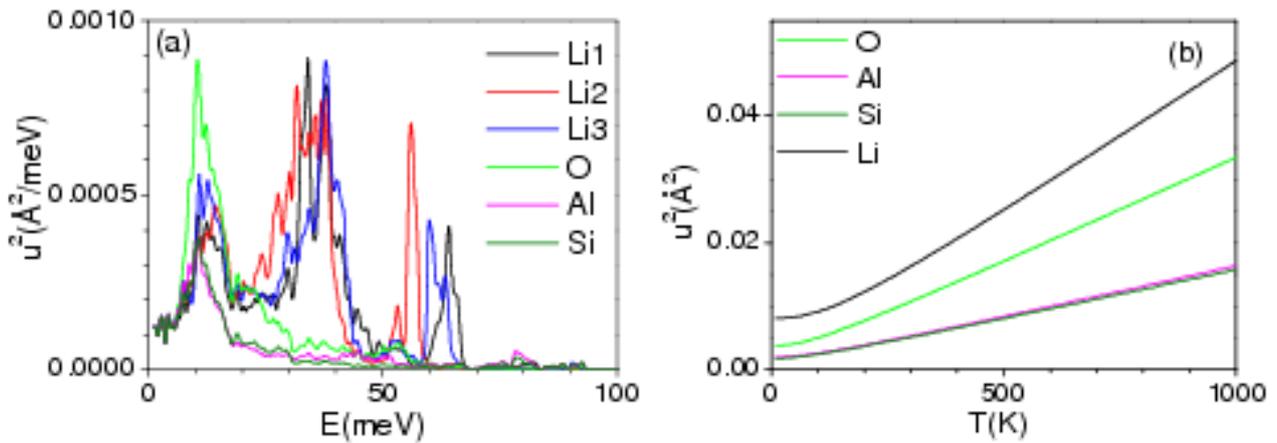

FIG. 7 (Color online) The calculated mean square amplitude of different atoms in the RT and HT phases of β-eucryptite as a function of temperature. The *xy* and *z* components of u² for an atom correspond to the in-plane and out-of-plane components, respectively.

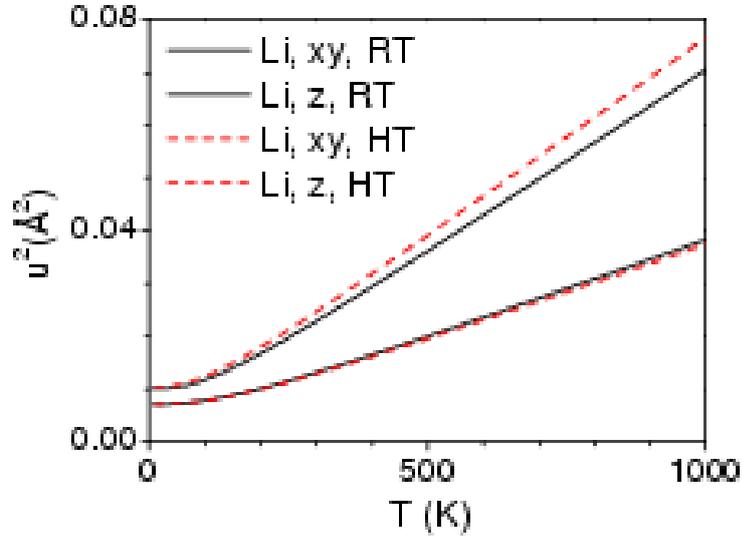

FIG. 8 (Color online) The calculated phonon dispersion curves for the HT-phase of β-eucryptite, at c lattice parameters of 11.476 Å and 11.337 Å, along the high symmetry directions in the Brillion zone of the hexagonal unit cell. The Bradley-Cracknell notation is used for the high-symmetry points: Γ (0,0,0), M(1/2,0,0), K(1/3,1/3,0), A (0, 0, 1/2), H (1/3,1/3,1/2) and L (1/2,0,1/2).

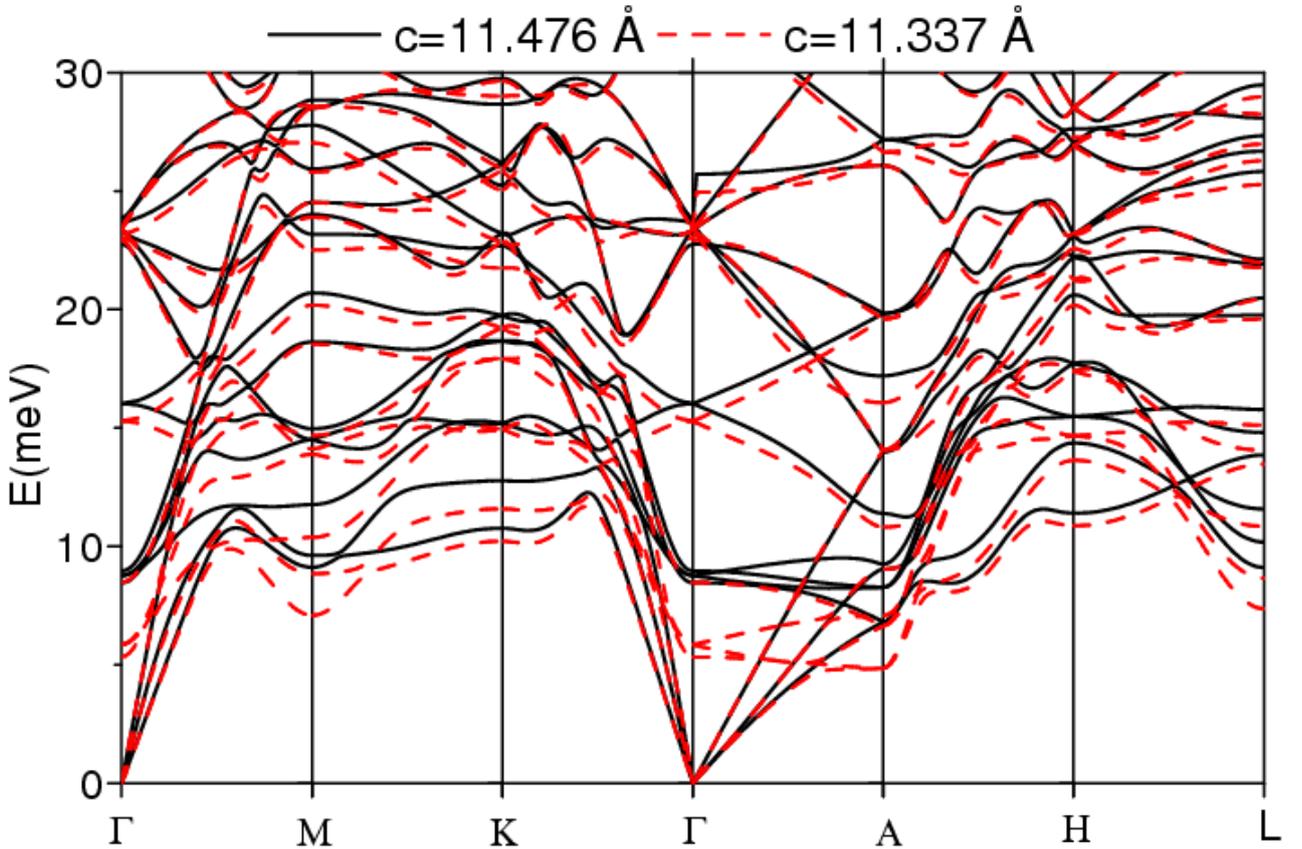

FIG. 9 (Color online): (a) the energy profile for the lithium A and S channels containing Li1 and (Li2 and Li3), respectively, in the RT-Phase, and HT-Phase of β-eucryptite. (b) Comparison between the barrier heights for the Li atom movement with and without Schottky vacancy in the HT-phase. The vacancy is initially located at a channel length parameter equal to c/3.

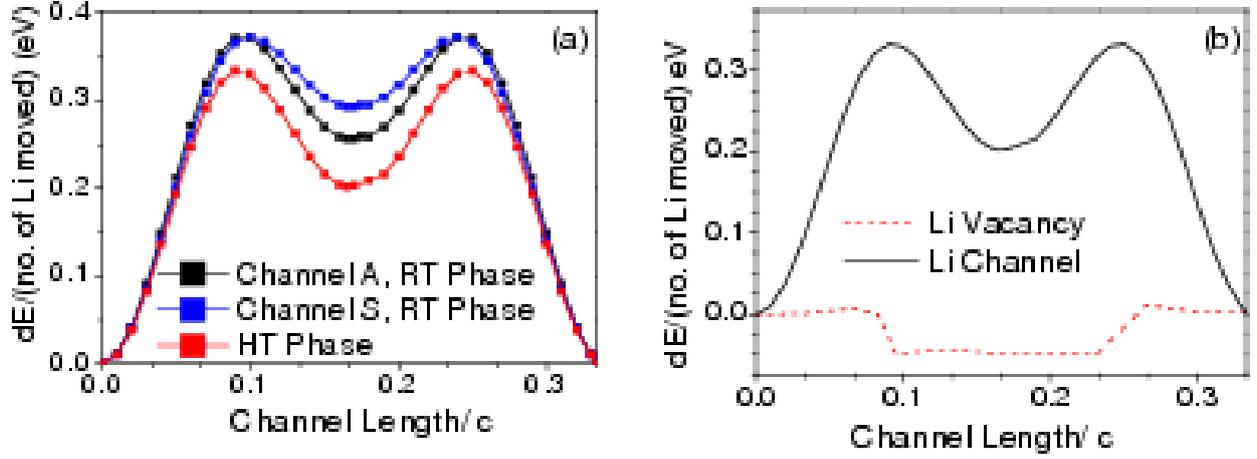

FIG. 10 (Color online) (a) The calculated Grüneisen parameters ($\Gamma_l, l=a, c$) as obtained from anisotropic stress along 'a' and 'c' axes. (b) Linear thermal expansion coefficients as a function of temperature along the a and c-axes. (c) The variation of the lattice parameters ($l=a$ or $c$) with temperature from ab-inito calculations and experiments [3]. (d) The contribution of phonons of energy $E$ to the linear thermal expansion coefficients ($\alpha_a$ and $\alpha_c$) as a function of $E$, at 300 K, in the room temperature phase of β-eucryptite.

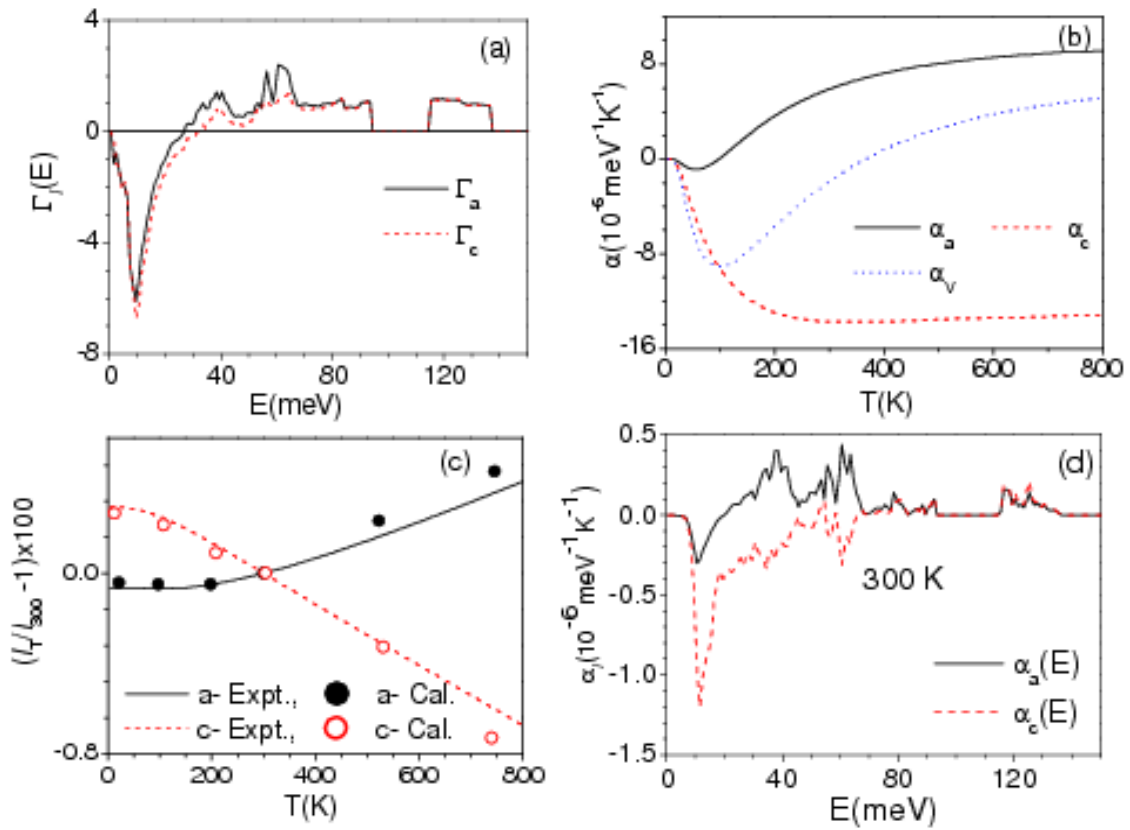

FIG. 11 (Color online): Phonon dispersion curves for the RT phase of β-eucryptite, at ambient (solid line) and 0.5 GPa (dash line) pressures, along the high symmetry directions in the Brillion zone of hexagonal unit cell. The Bradley-Cracknell notation is used for the high-symmetry points: Γ (0,0,0), M(1/2,0,0), K(1/3,1/3,0), A (0, 0, 1/2), H (1/3,1/3,1/2) and L (1/2,0,1/2).

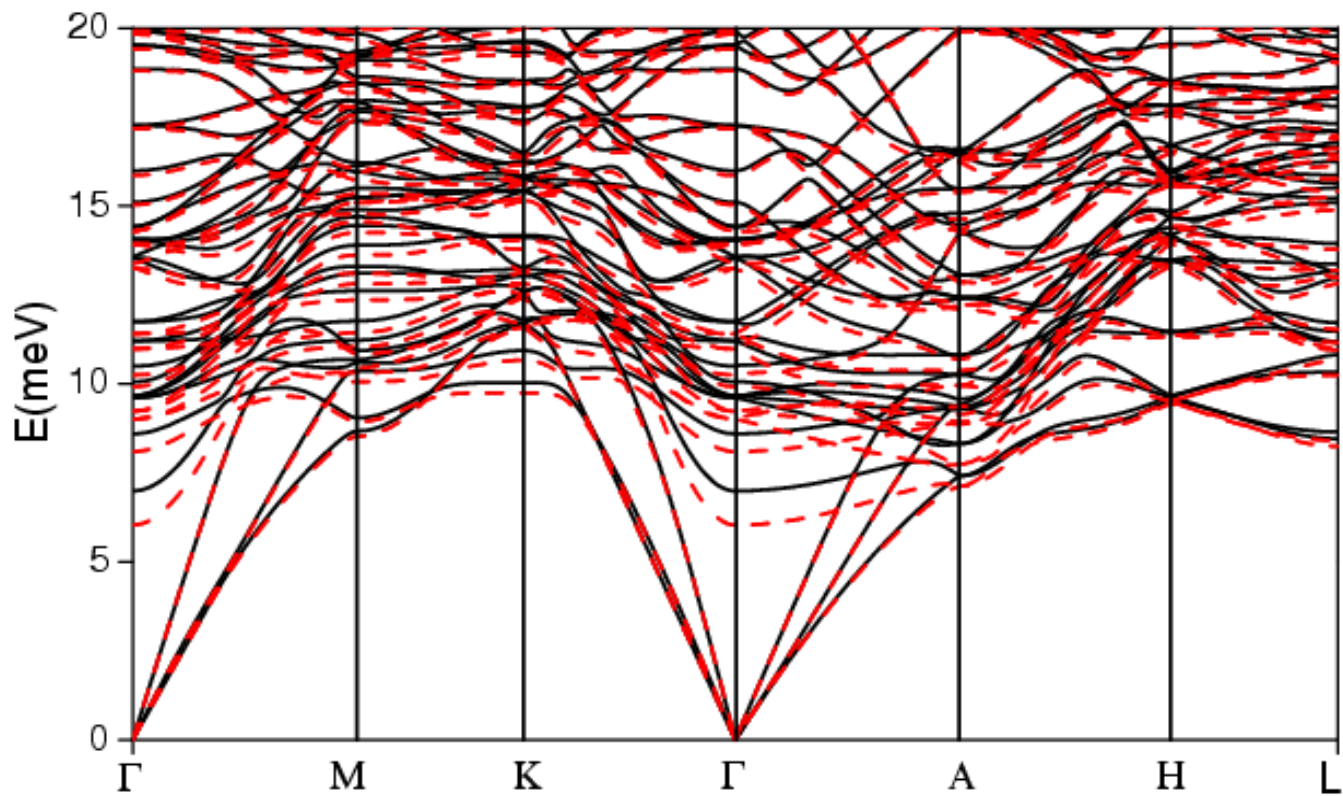

FIG. 12 (Color online) Schematic representation of the polarization vectors of selected zone centre optic phonon modes in β- eucryptite. The first, and second and third values specified below each plot indicate the phonon energy and Grüneisen parameters, respectively. The tetrahedral units around Al and Si are color-coded by blue and red respectively, while the Li atoms are represented by a green color.

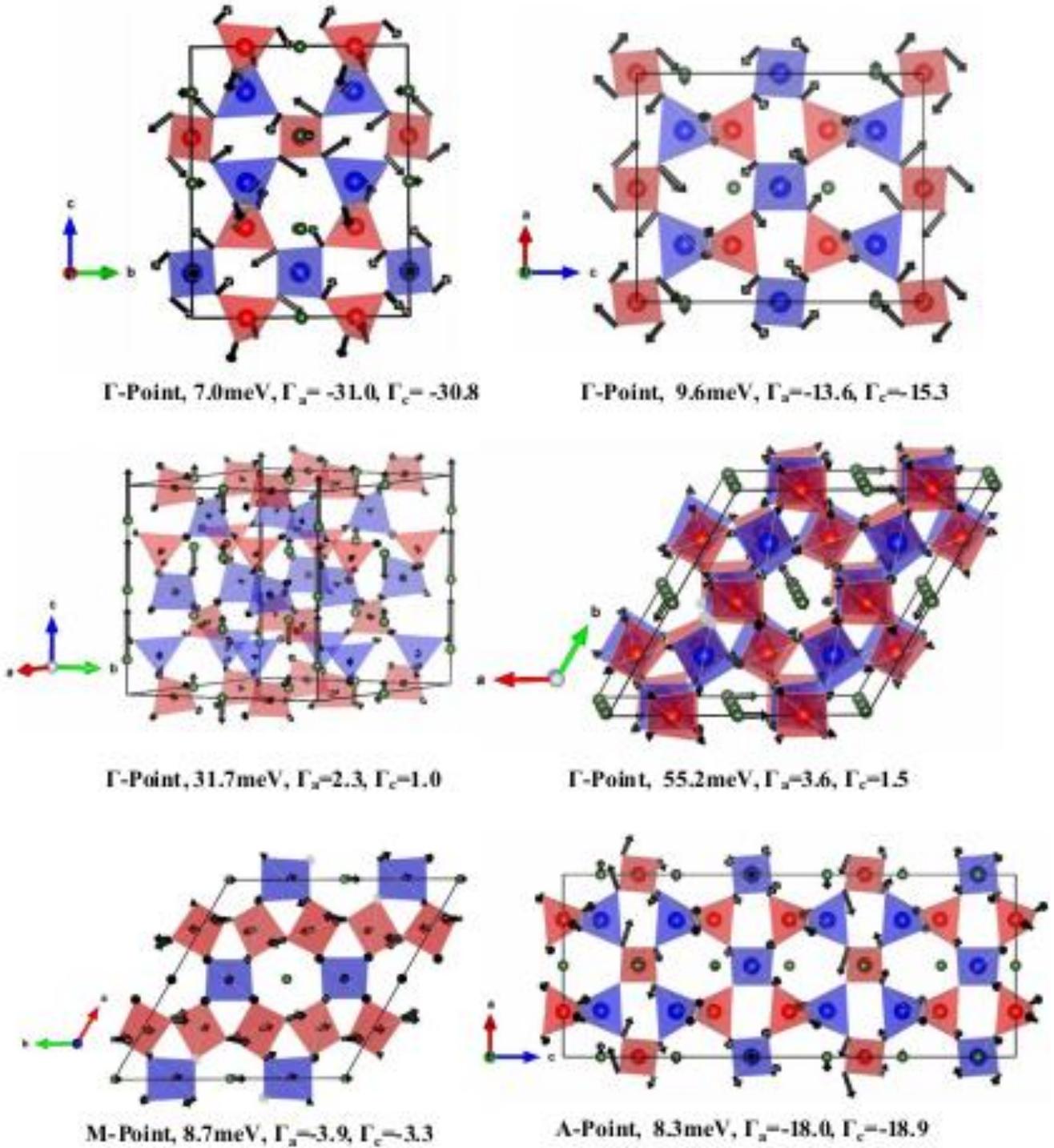

Γ-Point, 7.0meV, Γ$_a$= -31.0, Γ$_c$= -30.8

Γ-Point, 9.6meV, Γ$_a$=-13.6, Γ$_c$=-15.3

Γ-Point, 31.7meV, Γ$_a$=2.3, Γ$_c$=1.0

Γ-Point, 55.2meV, Γ$_a$=3.6, Γ$_c$=1.5

M-Point, 8.7meV, Γ$_a$=-3.9, Γ$_c$=-3.3

A-Point, 8.3meV, Γ$_a$=-18.0, Γ$_c$=-18.9

FIG. 13 (Color online) The calculated potential energy profile of the zone centre phonon mode at 7.0 meV, in the room temperature (RT) phase of β-eucryptite. The corresponding Grüneisen parameters are $\Gamma_a = -31.0$, $\Gamma_c = -30.8$.

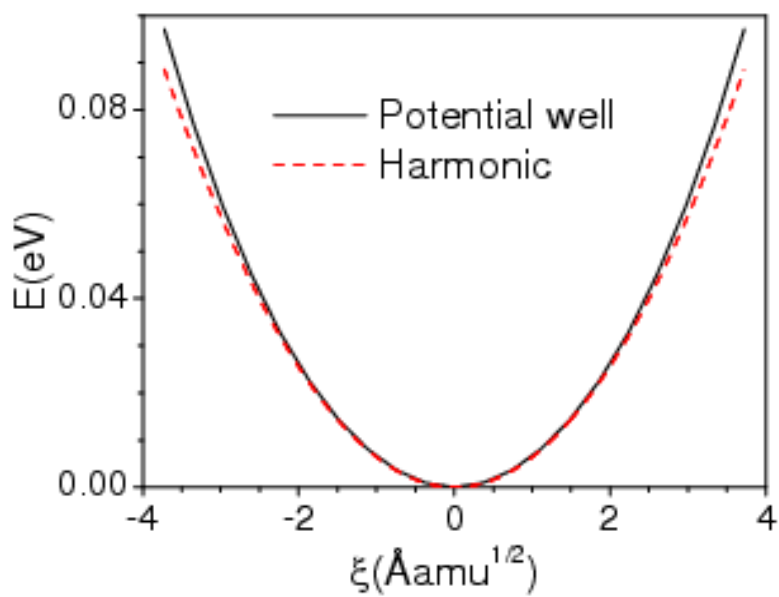

FIG. 14 (Color online) The temperature dependence of the zone centered phonon mode at 7.0meV. The mode has Grüneisen parameters of $\Gamma_a = -31.0$, $\Gamma_c = -30.8$.

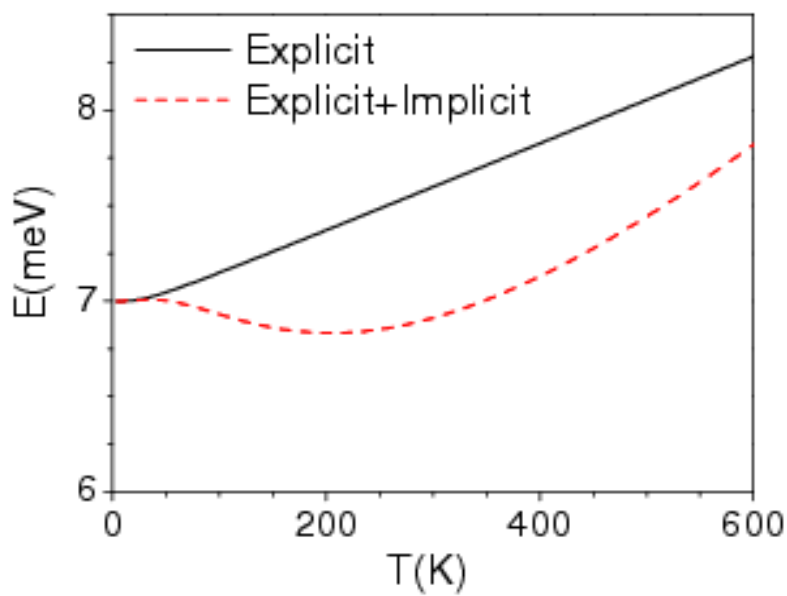

FIG. 15 (Color online) The calculated phonon dispersion curves for the room temperature β-eucryptite phase at 2.0 GPa along the high symmetry directions in the Brillion zone of the hexagonal unit cell. The Bradley-Cracknell notation is used for the high-symmetry points: Γ (0,0,0), M(1/2,0,0), K(1/3,1/3,0), A (0, 0, 1/2), H (1/3,1/3,1/2) and L (1/2,0,1/2).

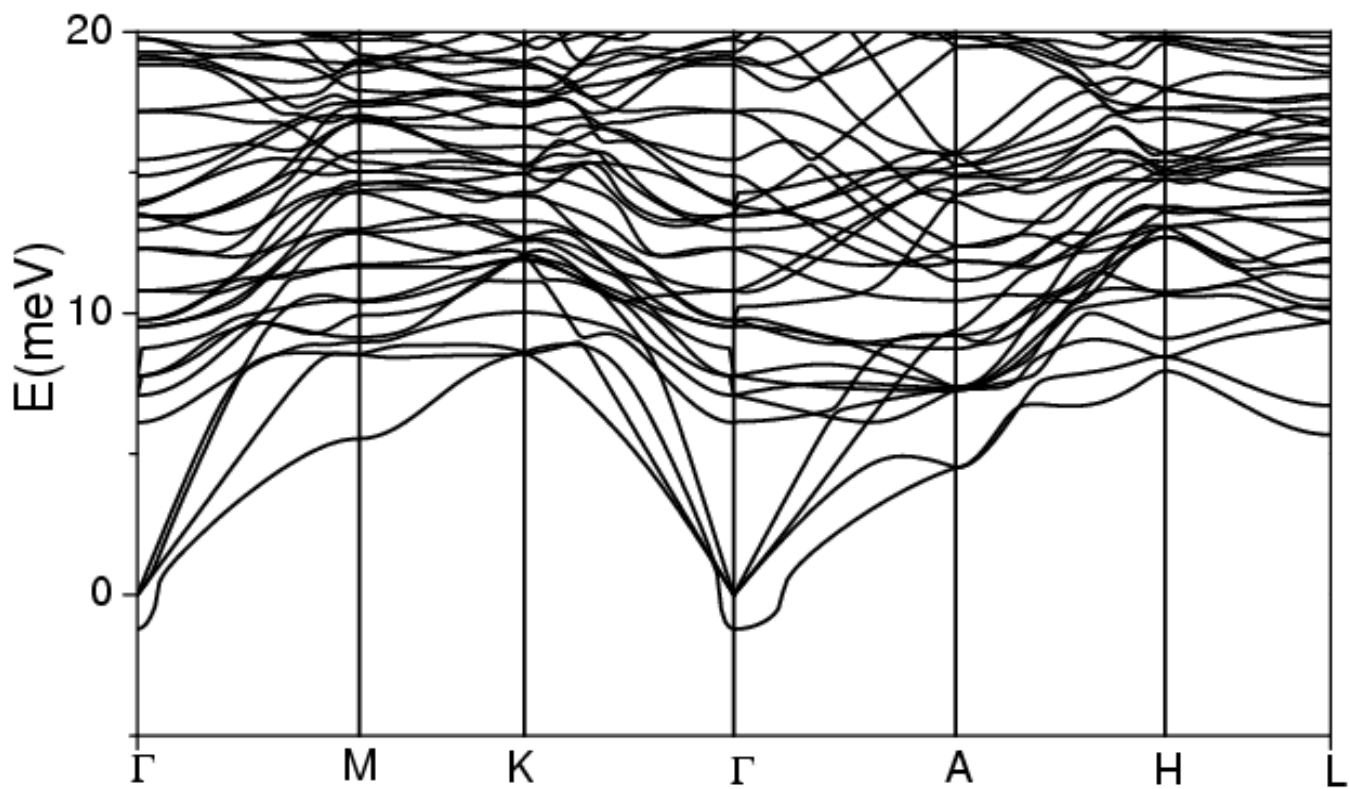